\documentclass[twocolumn,showpacs,amsmatsh,amssymb,aps,prc,groupedaddress,superscriptaddress,nofootinbib]{revtex4-1}

\usepackage[caption=false]{subfig}
\usepackage{tabularx}
\usepackage{graphicx}
\usepackage{amsmath}
\usepackage{dcolumn}
\usepackage{todonotes}
\usepackage{epsfig}
\usepackage{bm}
\usepackage{gensymb}
\usepackage{textcomp} 
\usepackage{hyperref}
\usepackage{color}

\begin{document}

\title{Impact of low-energy nuclear excitations on neutrino-nucleus scattering at MiniBooNE and T2K kinematics}

\author{V.~Pandey}
\email{Vishvas.Pandey@UGent.be}
\affiliation{Department of Physics and Astronomy, Ghent University,\\ 
Proeftuinstraat 86, B-9000 Gent, Belgium}
\author{N.~Jachowicz}
\affiliation{Department of Physics and Astronomy, Ghent University,\\
Proeftuinstraat 86, B-9000 Gent, Belgium}
\author{M.~Martini}
\affiliation{Department of Physics and Astronomy, Ghent University,\\
Proeftuinstraat 86, B-9000 Gent, Belgium}
\affiliation{ESNT, CEA, IRFU, Service de Physique Nucl\'eaire, \\
Universit\'e de Paris-Saclay, F-91191 Gif-sur-Yvette Cedex, France.}
\author{R.~Gonz\'{a}lez-Jim\'{e}nez}
\affiliation{Department of Physics and Astronomy, Ghent University,\\ 
Proeftuinstraat 86, B-9000 Gent, Belgium}
\author{J.~Ryckebusch}
\affiliation{Department of Physics and Astronomy, Ghent University,\\ 
Proeftuinstraat 86, B-9000 Gent, Belgium}
\author{T.~Van Cuyck}
\affiliation{Department of Physics and Astronomy, Ghent University,\\
Proeftuinstraat 86, B-9000 Gent, Belgium}
\author{N.~Van Dessel}
\affiliation{Department of Physics and Astronomy, Ghent University,\\ 
Proeftuinstraat 86, B-9000 Gent, Belgium}




\begin{abstract}
\begin{description}
\item[Background] Meticulous modeling of neutrino-nucleus interactions is essential to achieve the unprecedented precision goals of
present and future accelerator-based neutrino-oscillation experiments.
\item[Purpose] Confront our calculations of charged-current quasielastic cross section with the measurements of MiniBooNE and T2K, and 
to quantitatively investigate the role of nuclear-structure effects, in particular, low-energy nuclear excitations in forward muon scattering.
\item[Method] The model takes the mean-field (MF) approach as the starting point, and solves Hartree-Fock (HF)
equations using a Skyrme (SkE2) nucleon-nucleon interaction. Long-range nuclear correlations are taken into account
by means of the continuum random-phase approximation (CRPA) framework.
\item[Results] We present our calculations on flux-folded double differential, and flux-unfolded total cross sections off $^{12}$C and compare them with 
MiniBooNE and (off-axis) T2K measurements. We discuss the importance of low-energy nuclear excitations for the forward bins.  
\item[Conclusions] The HF and CRPA predictions describe the gross features of the measured cross sections. They underpredict the data 
(more in the neutrino than in the antineutrino case) because of the absence of processes beyond pure quasielastic scattering in our model. 
At very forward muon scattering, low-energy HF-CRPA nuclear excitations ($\omega < $ 50 MeV) account for 
nearly 50\% of the flux-folded cross section. This extra low-energy strength is a feature of the detailed microscopic nuclear model used here, 
that is not accessed in a Fermi-gas based approach.
\end{description}
\end{abstract}



\pacs{25.30.Pt, 13.15.+g, 24.10.Jv, 24.10.Cn}

\maketitle


\section{Introduction}

The study of neutrino oscillations is moving into an era of precision with an intense enhancement in the activities of accelerator-based
neutrino-oscillation experiments. Substantial progress has been made in the determination of the mass-squared differences and mixing angles. 
However, in order to improve the precision of the analysis, a rigorous 
description of neutrino-nucleus cross sections is required. The progress and issues related to the cross sections in this context were 
recently reviewed in Refs.~\cite{Review:2012, Review:2014, Mosel:2016}.
In recent years, several collaborations have reported muon neutrino cross sections on nuclei
~\cite{MiniBooNE:CCQE_Nu, MiniBooNE:NC_pi0, MiniBooNE:NCE_Nu, SciBooNE_1, MiniBooNE:CC_pi+, MiniBooNE:CC_pi0, ArgoNeuT_1,
MiniBooNE:CCQE_Anu, T2K:Inc_QE_nu, Minerva_1, Minerva_2, ArgoNeuT_2, T2K_2, Minerva_3, Minerva_4, ArgoNeuT_3, Minerva_5,
Minerva_6, MiniBooNE:NCE_Anu, T2K_3, T2K:CCQE_nu, T2K_4}.  
The challenges faced in these efforts, and especially those related to the neutrino-nucleus signal in the detector, need detailed 
microscopic 
neutrino-interaction models that can describe the variety of nuclear effects over the broad kinematical range probed. 
A thorough comparison of the cross-section measurements with theoretical predictions is crucial to assess the role of nuclear effects in the target's
response and to reduce the systematic uncertainties in the extraction of the oscillation parameters.
 
In this work, we aim at discussing the results of calculations for charged-current (CC) $\nu_\mu$ and
$\bar{\nu}_\mu$ scattering on $^{12}$C, at the kinematics of the MiniBooNE and T2K experiments. In particular, we focus on comparing
our calculations with charged-current quasi elastic (CCQE) $\nu_\mu$ and
$\bar{\nu}_\mu$ measurements of MiniBooNE~\cite{MiniBooNE:CCQE_Nu, MiniBooNE:CCQE_Anu}, and inclusive
and CCQE $\nu_\mu$ measurements of (off-axis near detector ND280) T2K~\cite{T2K:Inc_QE_nu, T2K:CCQE_nu}. 
One of the major objectives of this work is the investigation of the 
role of neutrino-induced low-energy nuclear collective excitations in MiniBooNE and T2K's signal. To this end we adopt a 
continuum random-phase approximation (CRPA) model.

The article is organized as follows. In Section~\ref{formalism}, we briefly discuss the main ingredients of our model. 
Sec.~\ref{results} is divided in three parts: We compare the flux-folded
double-differential CRPA cross sections with the measurements of MiniBooNE and T2K in Sec.~\ref{double-differential}. 
In order to asses the contributions stemming from low-energy nuclear excitations, we discuss the specific case of
forward muon scattering bins in Sec.~\ref{forward_scattering}. In Sec.~\ref{total_cross-section}, we show
flux-unfolded total cross sections. The conclusions are presented in Sec.~\ref{conclusions}.

\section{Formalism}\label{formalism}

The CRPA model was originally 
developed to describe exclusive electron- and photo-induced nucleon knockout reactions~\cite{Ryckebusch:1988, Ryckebusch:1989}. The model was 
later used to predict neutrino scattering at supernova energies both in charged-current (CC) and neutral-current (NC) 
reactions~\cite{Jachowicz:NC_1999, Jachowicz:CC_2002, Jachowicz:NC_2004, Jachowicz:SN_2006}. The formalism was further extended to the 
QE reaction region and
successfully tested against electron-scattering data for a variety of nuclear targets in the QE 
region~~\cite{Pandey:CCQE_Anu2014, Pandey:CRPA2014, Martini:2016}.
Here, we briefly summarize the essence of our model. The starting point of the description of the nuclear dynamics is a mean field (MF). 
We solve the Hartree-Fock (HF) equations using the Skyrme SkE2 nucleon-nucleon interaction~\cite{Ryckebusch:1989, Waroquier:1987}.
Once the bound and continuum single-nucleon wave functions are determined, long-range correlations are taken into account by means of a 
CRPA approach based on a Green's function formalism.
The CRPA describes an excited state as a linear combination of particle-hole
(1p1h) and hole-particle (1h1p) excitations out of a correlated nuclear ground state
\begin{equation}
 \arrowvert \Psi_{RPA}^{C} \rangle = \sum_{C'} \left[ X_{C, C^{'}} ~ \arrowvert p'h'^{-1} \rangle - 
 ~Y_{C, C^{'}}~ \arrowvert h'p'^{-1} \rangle \right]~,
\end{equation}
where $C$ represents the complete set of quantum numbers of an accessible single-nucleon knockout channel. The RPA polarization propagator $\Pi^{(RPA)}$
is obtained by the iteration of
the first order contributions to the particle-hole Green's function $\Pi^{(0)}$ and is obtained as the solution to the equation
\begin{eqnarray}
 & & \Pi^{(RPA)} (x_1,x_2;E_x) =  \Pi^{(0)} (x_1,x_2;E_x) \nonumber \\
&   &  + \frac{1}{\hbar} \int dx dx' \Pi^{0} (x_1,x;E_x)  
\tilde{V}(x, x') \Pi^{(RPA)} (x',x_2;E_x), \nonumber \\ && \label{propagator}
\end{eqnarray}
where $E_x$ is the excitation energy of the target nucleus and $x$ is a shorthand notation for the combination of the spatial, 
spin and isospin coordinates. The $\Pi^{(0)}$ in Eq.~(\ref{propagator}) corresponds 
to the HF contribution to the polarization propagator and $\tilde{V}$ denotes the antisymmetrized nucleon-nucleon SkE2
interaction.  

The SkE2 interaction was optimized against ground-state and low-excitation properties of 
spherical nuclei. Its strength lies in its ability to describe nuclear excitations in the few 10s of MeV energy range. 
The same SkE2 two-body interaction, that is used to solve the HF equations, is used to calculate the CRPA polarization
propagator. 
In order to restrain the SkE2 force from 
becoming unrealistically strong at high virtuality $Q^{2}$, a dipole hadronic form factor is introduced at the nucleon-nucleon interaction 
vertices~\cite{Pandey:CRPA2014}. 
The continuum wave functions are obtained by solving the 
positive-energy Schr\"odinger equation with appropriate boundary conditions. 
Hence, the distortion effects (escape width) from the residual nucleons on the outgoing nucleon is taken into account.
A folding procedure is used to take into account also the spreading width of the particle states~\cite{Pandey:CRPA2014}, which makes the description 
of giant resonances more realistic within the CRPA approach.
In order to consider the influence of the nuclear Coulomb field on the outgoing lepton,
a modified effective momentum approximation~\cite{Engel:1998} is used. Further, to improve our description at higher momentum transfers, we have implemented
relativistic kinematic corrections~\cite{Donnelly:1998}. The world-averaged axial mass value $M_{A}$ = 1.03 GeV was used
for all the calculations in this paper.


\section{Cross section analysis}\label{results}

\begin{figure}
\includegraphics[width=0.99\columnwidth]{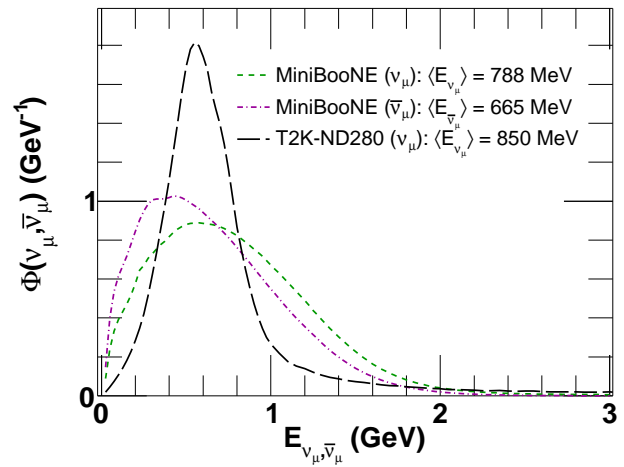}
\caption{(Color online) Normalized MiniBooNE $\nu_{\mu}$~\cite{MiniBooNE:CCQE_Nu}, $\bar{\nu}_{\mu}$~\cite{MiniBooNE:CCQE_Anu}
and T2K~\cite{T2K:Inc_QE_nu} (off-axis ND280) $\nu_{\mu}$ fluxes.}
\label{fig_flux}
\end{figure}

Both MiniBooNE and T2K use a target rich in $^{12}$C. Their fluxes~\cite{MiniBooNE:CCQE_Nu, MiniBooNE:CCQE_Anu, T2K:Inc_QE_nu} are slightly 
different, as shown in Fig~\ref{fig_flux}. Both $\nu_{\mu}$ beams have average energies around 800 MeV while the $\bar{\nu}_{\mu}$ MiniBooNE beam 
has a slightly lower average energy. The T2K beam is more sharply peaked, and receives less contributions beyond 1 GeV, than the
MiniBooNE one.

\subsection{Flux-folded double differential cross sections}\label{double-differential}

\begin{figure*}

\begin{minipage}{\dimexpr\linewidth-0.50cm\relax}
\rotatebox{90}{\mbox{\large \bm{$\textlangle d^{2}\sigma/dT_{\mu}dcos\theta_{\mu} \textrangle (10^{-42}cm^{2}MeV^{-1})$}}}
\vspace*{-8.5cm}\hspace*{16.8cm}
\end{minipage}%

\begin{minipage}{\dimexpr\linewidth-0.50cm\relax}
\includegraphics[width=0.94\textwidth]{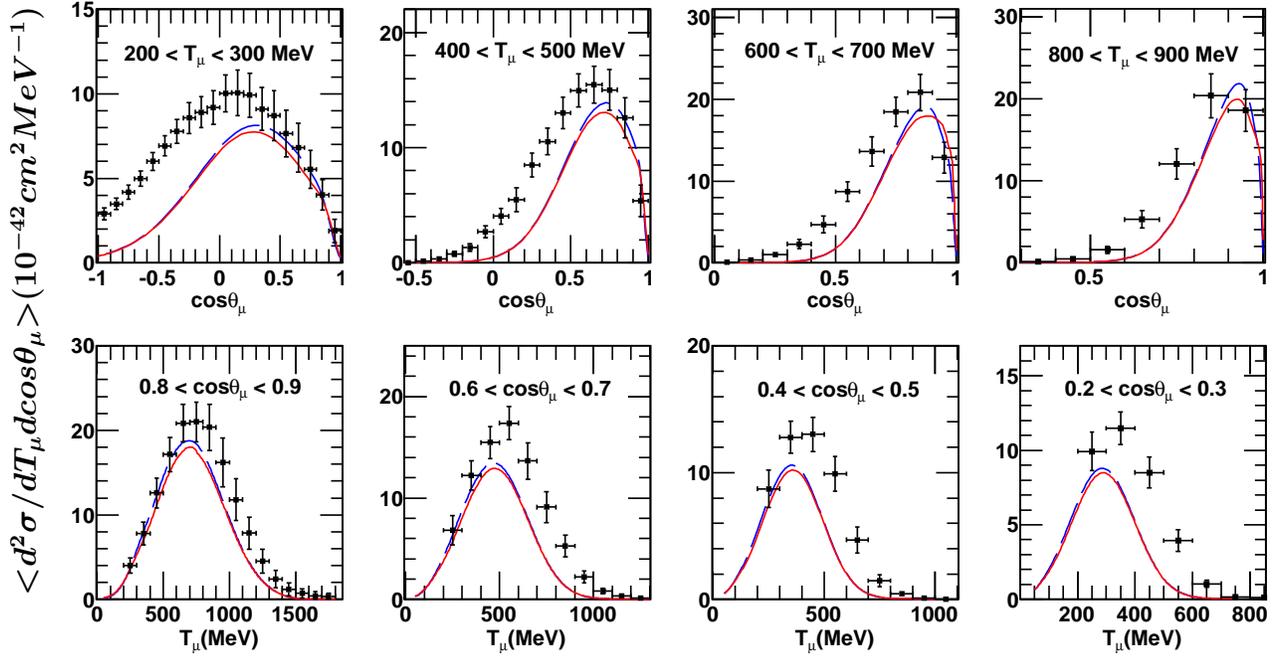}
\caption{(Color online) MiniBooNE flux-folded double-differential cross sections per target 
neutron for $^{12}$C$(\nu_{\mu},  \mu^{-})X$, plotted as a function of $\cos\theta_{\mu}$ for 
different $T_{\mu}$ values (top) and as a function of $T_{\mu}$ for different ranges of 
$\cos\theta_{\mu}$ (bottom). Solid curves are CRPA and dashed curves are HF results. 
MiniBooNE data including shape uncertainties are taken from Ref.~\cite{MiniBooNE:CCQE_Nu}.
The data contains an additional normalization uncertainty 
of 10.7\%, not included here.}
\label{fig_ddxs_nu_MiniBooNE_folded}
\end{minipage}
\end{figure*}

\begin{figure*}

\begin{minipage}{\dimexpr\linewidth-0.50cm\relax}
\rotatebox{90}{\mbox{\large \bm{$\textlangle d^{2}\sigma/dT_{\mu}dcos\theta_{\mu} \textrangle (10^{-42}cm^{2}MeV^{-1})$}}}
\vspace*{-8.5cm}\hspace*{16.8cm}
\end{minipage}%

\begin{minipage}{\dimexpr\linewidth-0.50cm\relax}
\includegraphics[width=0.94\textwidth]{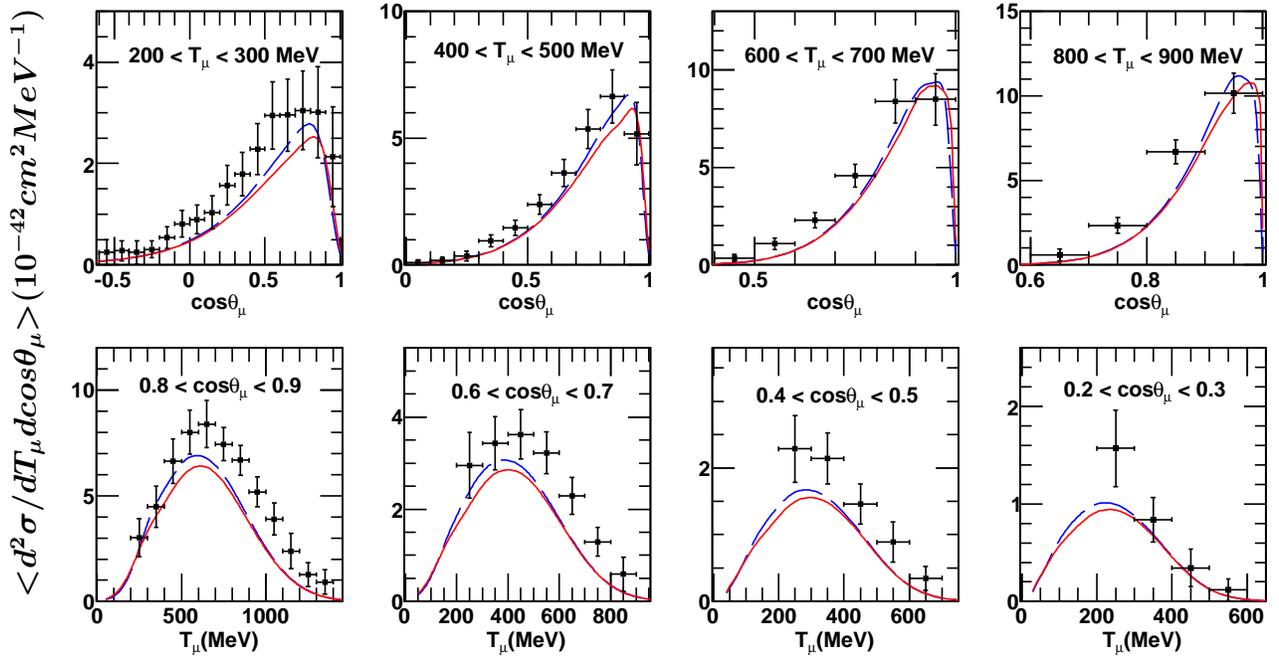}
\caption{(Color online) Same as Fig.~\ref{fig_ddxs_nu_MiniBooNE_folded} but for the process 
$^{12}$C$(\bar{\nu}_{\mu},  \mu^{+})X$. Solid curves are CRPA and dashed curves are HF calculations. 
MiniBooNE data including shape uncertainties are taken from Ref.~\cite{MiniBooNE:CCQE_Anu}. 
The data contain an additional normalization uncertainty of 17.4\%, not included here.}
\label{fig_ddxs_anu_MiniBooNE_folded}
\end{minipage}
\end{figure*}

\begin{figure}
\includegraphics[width=0.99\columnwidth]{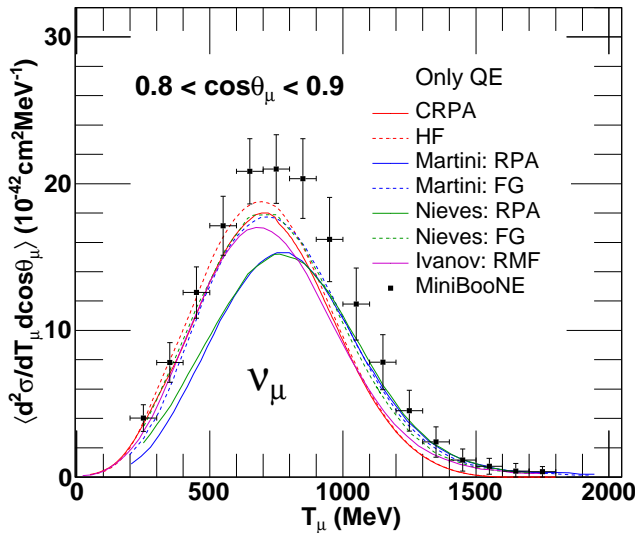}
\caption{(Color online) MiniBooNE flux-folded cross section per target
  neutron for $^{12}$C$(\nu_{\mu}, \mu^{-})X$ at $0.8 <
  \cos\theta_{\mu} < 0.9$. The CRPA and HF predictions are compared with those of Martini {\it et al.} 
  \cite{Martini:2011}, Nieves {\it et al.}~\cite{Nieves:2012} and Ivanov {\it et al.}~\cite{Ivanov:2013}.}
\label{fig_MiniBooNE_QE_Comparison}  
\end{figure}

\begin{figure}
\includegraphics[width=0.99\columnwidth]{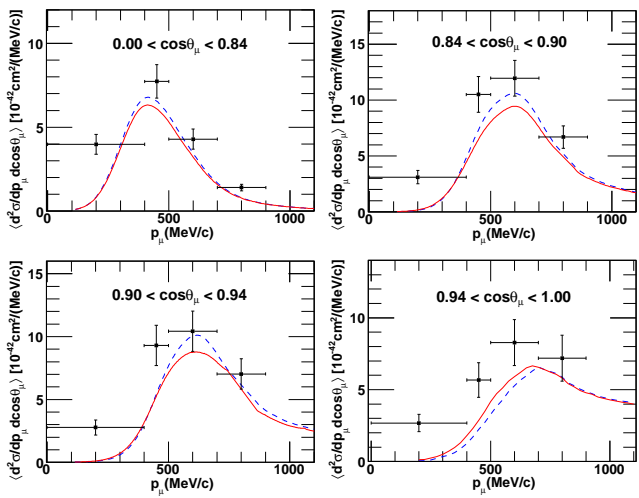}
\caption{(Color online) T2K flux-folded inclusive CC double-differential cross sections per target 
nucleon on $^{12}$C plotted as a function of muon momentum $p_{\mu}$, for different bins of 
$\cos\theta_{\mu}$. CRPA (solid curves) and HF (dashed-curves) are compared with T2K measurements
of~\cite{T2K:Inc_QE_nu}.}
\label{fig_ddxs_nu_T2K_folded}
\end{figure}


\begin{figure*}
\includegraphics[width=0.90\textwidth]{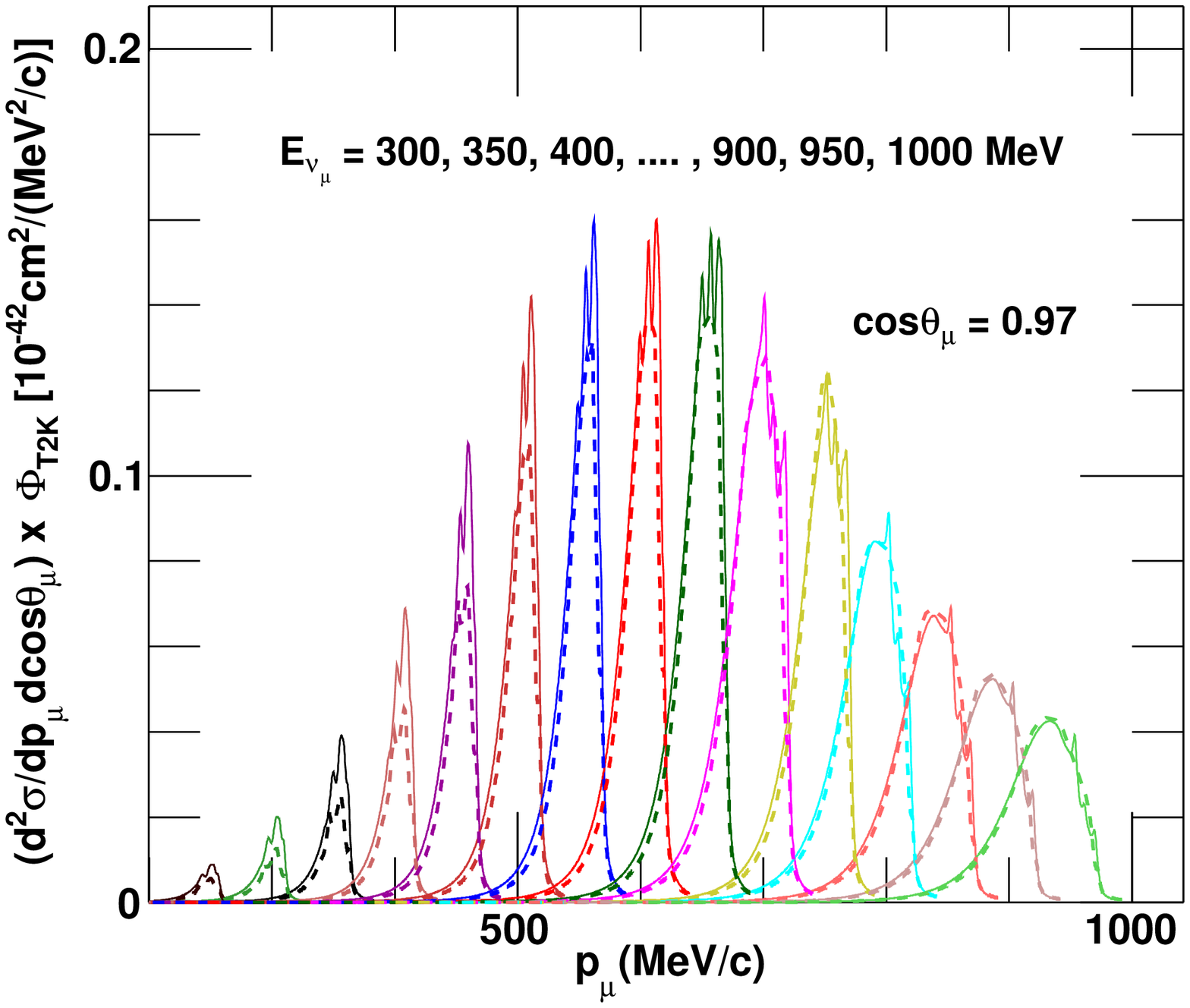}
\caption{(Color online) Double differential CRPA (solid curves) and HF (dotted curves) cross sections on $^{12}$C for $\cos\theta_\mu$ = 0.97
and for fixed neutrino energies from 300 MeV to 1000 MeV, weighted with the T2K $\nu_\mu$ flux (of Fig.~\ref{fig_flux})
and plotted as a function of $p_{\mu}$.}
\label{fig_ddxs_nu_T2K_Cos97}
\end{figure*}

We present CC pure QE neutrino cross sections folded with the MiniBooNE flux in Fig.~\ref{fig_ddxs_nu_MiniBooNE_folded}. The top 
panels are plotted as a function of the muon scattering angle $\cos\theta_{\mu}$ for several bins of muon kinetic energies $T_{\mu}$ 
and the bottom panels are plotted as function of $T_{\mu}$ for different ranges of $\cos\theta_{\mu}$. The calculated cross sections are 
averaged over the $T_{\mu}$ and $\cos\theta_{\mu}$ ranges. We compare HF and CRPA calculations with the experimental 
data of MiniBooNE~\cite{MiniBooNE:CCQE_Nu}. The HF and CRPA calculations reproduce the shape of the measured cross sections. 
In the top panels, the CRPA cross sections are slightly higher
than the HF ones for $\cos\theta_{\mu}$ approaching 1, owing to extra contributions stemming from low-energy excitations. 
For forward scattering the 1p-1h CRPA model reasonably reproduces the data, whereas it tends to underestimate the measured cross sections at backward scattering.
The measurement of CCQE neutrino~\cite{MiniBooNE:CCQE_Nu} and 
antineutrino~\cite{MiniBooNE:CCQE_Anu} cross sections by the MiniBooNE collaboration sparked off discussions about the nuclear 
effects active in the broad energy range covered by the flux. The CCQE(-like) cross section in MiniBooNE is defined
as the process where one muon and no pions are observed in the final state.
Corrections to genuine QE processes stem from multi-nucleon correlations in the target nuclei. Those multi-nucleon processes (like meson-exchange currents
(MEC), $\Delta$-isobar currents and short-range correlations) give rise to additional sources of strength in the nuclear response: 
a correction in the
single-nucleon knockout channel, and a non-vanishing strength in multinucleon knockout channel.
The necessity to include multinucleon effects to successfully describe the CCQE MiniBooNE data has been suggested~\cite{Martini:2009, Martini:2010}
and confirmed 
by several independent models~\cite{Amaro:2011, Nieves:2011, Bodek:2011, Martini:2011, Nieves:2012, 
Amaro:2012, Lalakulich:2012, Nieves:2013, Martini:2013, Megias:2015, Tom:2016}. 
As expected, the exclusion of 
multi-nucleon channels in this work, results in an underestimation of the data. 

In Fig.~\ref{fig_ddxs_anu_MiniBooNE_folded}, we compare our 
flux-folded predictions for antineutrino cross sections with the MiniBooNE measurements of
Ref.~\cite{MiniBooNE:CCQE_Anu}. In this case, the CRPA predictions are closer to 
the MiniBooNE data than those for the neutrino calculations. This again confirms that the relative role of multi-nucleon excitations 
is more important for the neutrino than for the antineutrino cross sections, as discussed in Ref.~\cite{Martini:2010}. 

In Fig.~\ref{fig_MiniBooNE_QE_Comparison}, we present a QE-only comparison of our HF and CRPA calculations with 
the predictions of Martini {\it et al.}~\cite{Martini:2011}, Nieves {\it et al.}~\cite{Nieves:2012}, and Ivanov {\it et al.}~\cite{Ivanov:2013}. 
The comparison is presented for MiniBooNE flux-folded cross-sections off $^{12}$C at $0.8 < \cos\theta_{\mu} < 0.9$.
In the Martini {\it et al.} and Nieves {\it et al.} approaches, the QE predictions of both models (FG and RPA) almost coincide. 
There is a sizable RPA quenching in the predictions of both Martini {\it et al.}, and Nieves {\it et al.}.
The size of the quenching is smaller in the CRPA, resulting in a larger predicted cross section for the QE process.
The authors of Ref.~\cite{Martini:2011, Nieves:2012} attribute the
strong quenching in their model to the explicit inclusion of the Ericson-Ericson-Lorentz-Lorentz effect, which accounts for the possibility of a $\Delta$-hole excitation
in the RPA chain.
For a more detailed comparison of our HF-CRPA model with the model of Martini {\it et al.}, we refer the reader to Ref.~\cite{Martini:2016}. 
The relativistic mean-field (RMF) predictions of Ivanov {\it et al.} are lower than our HF ones around the peak, but the RMF generates more strength at the 
high-T$_{\mu}$ end. The pure QE RPA results of Martini {\it et al.} and Nieves {\it et al.} are significantly different from
HF, CRPA and RMF results. These difference can be assigned to the use of a detailed microscopic nuclear model in the HF and RMF calculations
compared to the FG ones.
Note that, the additional contribution from np-nh in Martini {\it et al.} and Nieves {\it et al.}, and from meson-exchange
current (MEC) in Ref.~\cite{Megias:2016} were included in these collaborations to describe the MiniBooNE data. These additional channels are not shown in
Fig.~\ref{fig_MiniBooNE_QE_Comparison} as we focused only on the pure QE channel. Still, one should be aware that the separation into different channels can be 
strongly model-dependent.
\begin{figure}
\includegraphics[width=0.99\columnwidth]{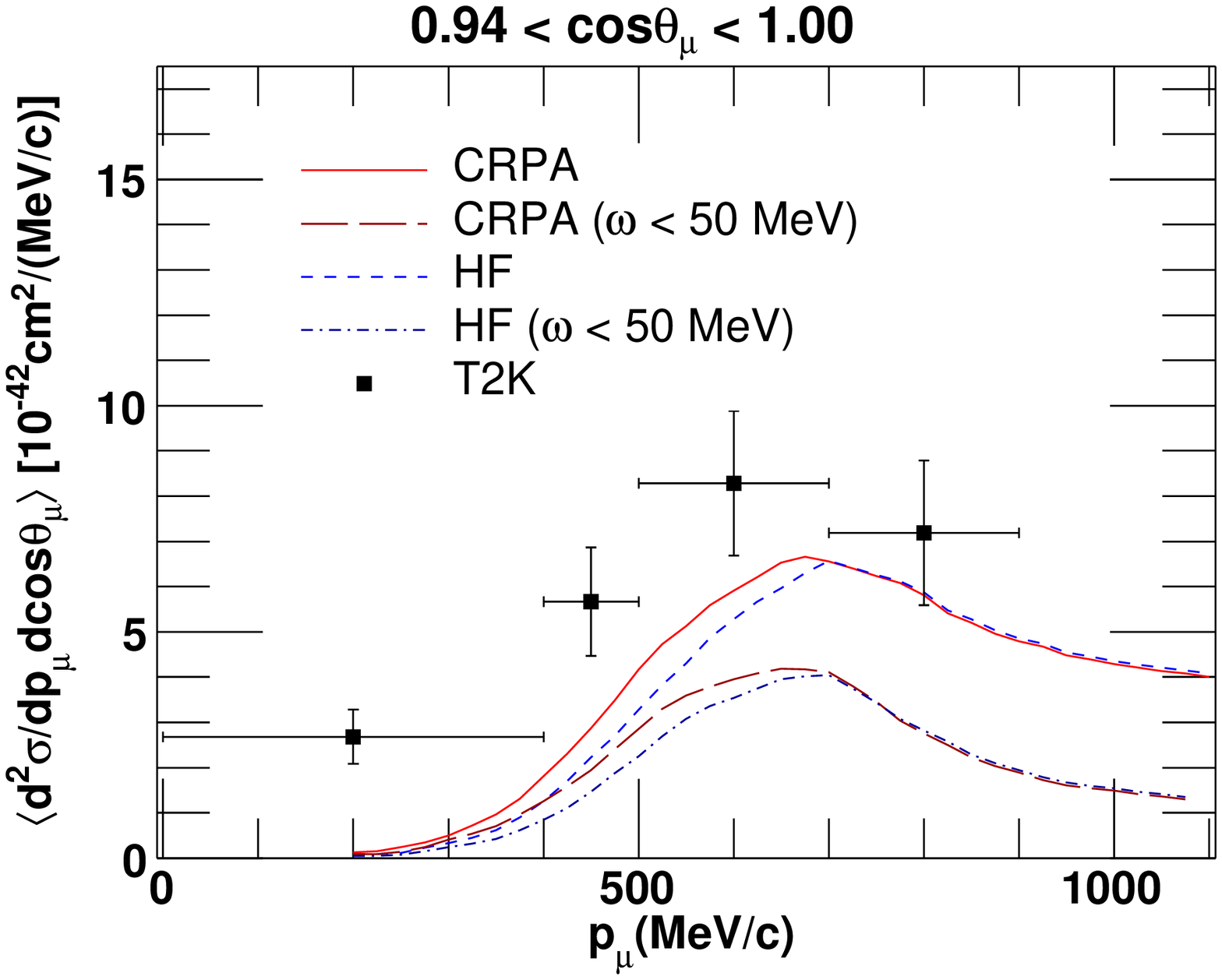}
\caption{(Color online) The most forward bin,  $ 0.94 < \cos\theta_{\mu} < 1.0$, of 
Fig~\ref{fig_ddxs_nu_T2K_folded}. T2K data are taken from Ref.~\cite{T2K:Inc_QE_nu}.}
\label{fig_ddxs_nu_T2K_folded_ForwardBin}
\end{figure}

\begin{figure}
\includegraphics[width=0.99\columnwidth]{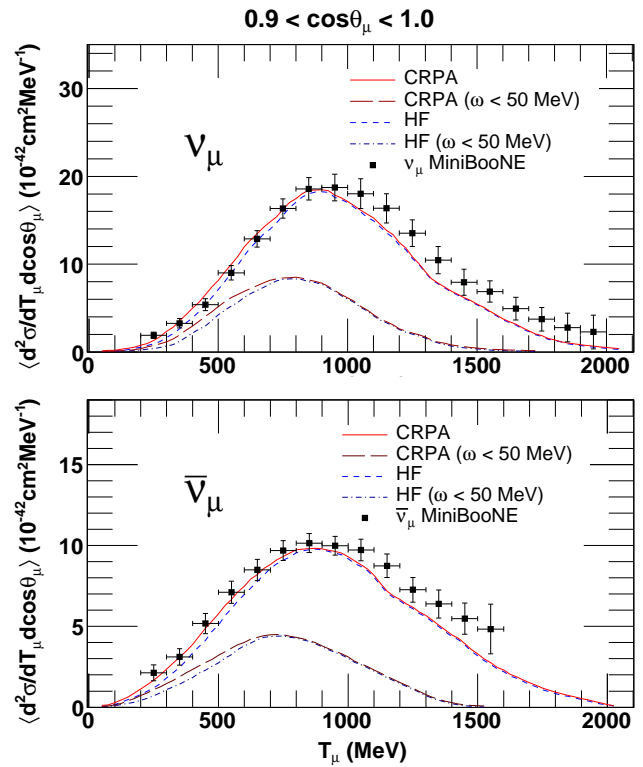}
\caption{(Color online) The $T_{\mu}$ dependence of the CCQE double-differential cross sections per target neutron folded with
MiniBooNE flux, for $ 0.9 < \cos\theta_{\mu} < 1.0$.
CRPA calculations are compared with MiniBooNE data of 
Ref.~\cite{MiniBooNE:CCQE_Nu}. Experimental error bars represent the shape uncertainties.}
\label{fig_ddxs_nu_MiniBooNE_folded_ForwardBin}
\end{figure}

\begin{figure}
\includegraphics[width=0.99\columnwidth]{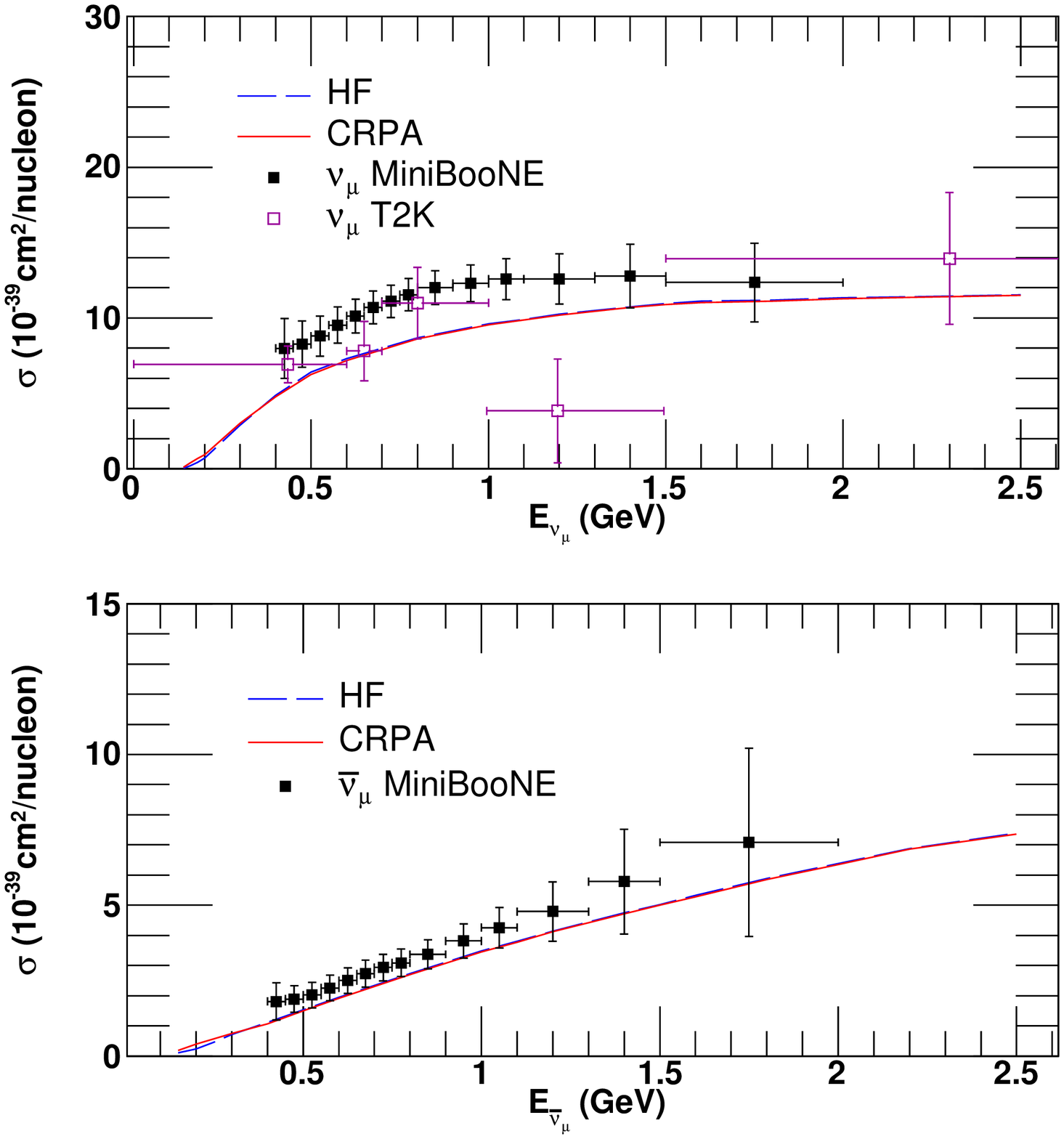}
\caption{(Color online) Total CCQE $^{12}$C$(\nu_{\mu},  \mu^{-})X$ and $^{12}$C$(\bar{\nu}_{\mu},  \mu^{+})X$
cross sections per target nucleon
plotted as a function of (anti)neutrino energy. The experimental data are taken from 
MiniBooNE ($\nu_\mu$)~\cite{MiniBooNE:CCQE_Nu}, T2K ($\nu_\mu$)~\cite{T2K:CCQE_nu} and MiniBooNE ($\bar\nu_\mu$)~\cite{MiniBooNE:CCQE_Anu}.}
\label{fig_xs_nu_MiniBooNE_T2K}
\end{figure}

The T2K collaboration reported on CC-inclusive double-differential cross sections as a function of muon momentum $p_{\mu}$ and scattering angle 
$\cos\theta_{\mu}$~\cite{T2K:Inc_QE_nu}, and CCQE total cross sections as a function of incident neutrino energies~\cite{T2K:CCQE_nu}. 
Ref.~\cite{Martini:2014} finds a 
satisfactory agreement with the T2K data, after inclusion of multinucleon and single-pion production channels. On the other hand, 
the relativistic Green's function (RGF) approach of Ref.~\cite{Meucci:2015}, which successfully describes the MiniBooNE data, 
underestimates the T2K results. Another comparison is presented in the superscaling approach of Ref.~\cite{Ivanov:2015}. 

We have computed the T2K $\nu_{\mu}$ flux-folded QE double-differential cross sections.
Our HF and CRPA results are confronted with the data in Fig.~\ref{fig_ddxs_nu_T2K_folded}. The cross sections are averaged over 
each $\cos\theta_{\mu}$ bin. CRPA cross sections reproduce the gross feature of the T2K data, but underestimate the data, as can be expected 
in absence of effects beyond QE. The underestimation is more pronounced for smaller values of $p_{\mu}$, which 
corresponds to the higher excitation energies where the inelastic channels beyond QE can be expected to have substantial contributions. 
For the most forward bin (0.94$~< \cos\theta_\mu <$ 1.0), the CRPA cross section 
is higher than the HF one for $p_{\mu} \lesssim $ 700 MeV/c. This behavior can be attributed to giant-resonances contributing a portion of the
CRPA strength.

\subsection{Forward scattering cross section}\label{forward_scattering}

In Ref.~\cite{Pandey:CRPA2014}, we stressed the importance of low-energy nuclear excitations for the forward muon scattering events
in MiniBooNE and T2K. Here we compare the most 
forward bin of the MiniBooNE and T2K data sets to explore the contributions emerging from low-energy excitations in these experiments. 
A substantial amount of the cross section strength 
in this kinematic region, where the excitation energy of the nucleus is $\lesssim$ 50 MeV, arises from collective nuclear excitations. 
As we have shown in Fig.~14 of Ref.~\cite{Pandey:CRPA2014} and Fig.~9 of Ref.~\cite{Martini:2016}, at these kinematics the longitudinal response 
generates major strength or comparable strength (depending on the neutrino energy) of the cross section with respect to the transverse response. 
Models that do not include collective effects can be expected to underestimate the data at
small scattering angles. 
The RGF predictions for T2K~\cite{Meucci:2015} significantly underestimate the 
data for 0.94$~< \cos\theta_{\mu} <~$1. In Ref.~\cite{Martini:2014}, even after the inclusion of multinucleon and one-pion production 
channels, which reproduced the data successfully in other angular bins, the prediction lacks strength in the forward bin. 
In Fig.~\ref{fig_ddxs_nu_T2K_Cos97}, we show HF and CRPA cross sections weighted with the T2K flux, for fixed neutrino energies
from 300 MeV to 1000 MeV and for fixed scattering angle  
$\cos\theta_\mu = $ 0.97. At these kinematics, the low-energy excitations significantly contribute to the cross section. In fact, the contributions
from $\omega <$ 50 MeV processes constitute a large part of the cross section for most incoming neutrino energies contributing to this forward 
angular bin. As expected, the low-energy excitation peaks are more pronounced in CRPA calculation than in HF ones
even up to incoming energies of $E_{\nu} =$ 1000 MeV. Though, the total integrated strength of both HF and CRPA
are almost same for $E_{\nu} \gtrsim$ 700 MeV.
Flux-folding of these cross sections washes out the low-energy excitation peaks and smooths the overall curve. Still, the important
low energy strength remains.
In Fig.~\ref{fig_ddxs_nu_T2K_folded_ForwardBin}, we show the contribution emerging from low-energy excitations
($\omega <$ 50 MeV) for the most forward bin of Fig.~\ref{fig_ddxs_nu_T2K_folded}.
This strength accounts for nearly 50\% of the cross section in this kinematic bin, representing lepton scattering angles up to 20$\degree$.
The relativistic Fermi-gas (RFG) based models, implemented in Monte-Carlo generators, in principle are not suitable to provide a detailed 
description of this kinematic range.
For the sake of completeness, in Fig.~\ref{fig_ddxs_nu_MiniBooNE_folded_ForwardBin} we compare our flux-folded results for  
0.90$~< \cos\theta_{\mu} <~$1 with the MiniBooNE measurements and separately plot the contribution from $\omega <$ 50 MeV.
This hints at the importance of an accurate description of neutrino-induced low-energy nuclear excitations in the most 
forward MiniBooNE and T2K measurements.

\subsection{Total cross section}\label{total_cross-section}
 
In Fig~\ref{fig_xs_nu_MiniBooNE_T2K}, we compare the computed CCQE $^{12}$C$(\nu_{\mu},  \mu^{-})X$ and $^{12}$C$(\bar{\nu}_{\mu},  \mu^{+})X$ 
total cross-section with
the data of MiniBooNE~\cite{MiniBooNE:CCQE_Nu, MiniBooNE:CCQE_Anu} and T2K~\cite{T2K:CCQE_nu}. Unlike 
the double-differential ones, the total experimental cross sections are model dependent as they are expressed as a function of
reconstructed energy~\cite{Martini:2012_2, Martini:2013_2, Nieves:2012_2, Lalakulich:2012_2, Leitner}, while the theoretical results as a function of true energy.
On average, the strength of the MiniBooNE measurements is higher than the T2K \textquoteleft QE-like\textquoteright~one. The measurements of 
these two data sets are quite comparable except for $E_{\nu_{\mu}} \simeq $ 1.2 GeV. The CRPA calculations are within the error bar of the T2K data,
but underpredict the MiniBooNE ones. The CRPA results agree much better with the antineutrino measurement of MiniBooNE. 
The HF and CRPA cross sections in both the neutrino and antineutrino case are almost coinciding with each other except
for E $<$ 250 MeV where CRPA cross section is higher than the HF one. 


\section{Conclusions}\label{conclusions}

We have calculated $\nu_{\mu}$-$^{12}$C and $\bar{\nu}_{\mu}$-$^{12}$C cross sections in kinematics corresponding with the MiniBooNE and T2K experiments. 
We compared flux-folded double differential cross sections with CCQE $\nu_\mu$ and $\bar\nu_{\mu}$ MiniBooNE measurements, and with inclusive T2K 
(off-axis) measurements. The CRPA cross sections 
compare favorably to the shape but underestimate the MiniBooNE data for backward muon scattering angles. 
The missing strength can be associated 
with the contribution from multi-nucleon knockout and single-pion production processes.
Still, a comparison of the flux-folded cross sections of MiniBooNE and T2K, shows that
for forward muon scattering, the neutrino-induced low-energy nuclear excitations ($\omega < $ 50 MeV) account for nearly 50\% of the
flux-folded cross section. 
These contributions, inaccessible in RFG-based Monte-Carlo generators, make a strong case 
for a more careful modeling of the forward signal in MiniBooNE and T2K-like experiments.



\acknowledgments
This work was supported by the Interuniversity Attraction Poles Programme initiated by the Belgian Science Policy Office (BriX network P7/12) and
the Research
Foundation Flanders (FWO-Flanders). M.M. acknowledges also the support and the framework 
of the ``Espace de Structure et de r\'eactions Nucl\'eaire Th\'eorique'' (ESNT, \url{http://esnt.cea.fr} ) at CEA.

\end{document}